\documentclass[preprint,secnumarabi,superscriptaddress,amssymb,nobibnotes,prc,preprintnumbers]{revtex4}

\usepackage{epsfig}
\usepackage{amsmath}
\usepackage{graphicx}
\usepackage{hyperref}

\begin{document}

\title{Next-to-leading order QCD predictions for dijet photoproduction in lepton-nucleus scattering at the future EIC
and at possible LHeC, HE-LHeC, and FCC facilities}

\author{V. Guzey}

\affiliation{National Research Center ``Kurchatov Institute'', Petersburg
 Nuclear Physics Institute (PNPI), Gatchina, 188300, Russia}
 
\author{M. Klasen}

\affiliation{Institut f\"ur Theoretische Physik, Westf\"alische
 Wilhelms-Universit\"at M\"unster, Wilhelm-Klemm-Stra{\ss}e 9, 48149
 M\"unster, Germany}




\begin{abstract}
  We calculate cross sections for inclusive dijet photoproduction in electron-nucleus
  scattering in the kinematics of the future EIC and the possible LHeC, HE-LHeC, and the FCC
  using next-to-leading order (NLO) perturbative QCD and nCTEQ15 and EPPS16 nuclear parton
  density functions (nPDFs). We make predictions for distributions in the dijet average
  transverse momentum ${\bar p}_T$, the average rapidity $\bar{\eta}$, the observed nuclear
  momentum fraction $x_A^{\rm obs}$, and the observed photon momentum fraction
  $x_{\gamma}^{\rm obs}$. Comparing the kinematic reaches of the four colliders, we find
  that an increase of the collision energy from the EIC to the LHeC and beyond extends the
  coverage in all four considered variables. Notably, the LHeC and HE-LHeC will allow one
  to probe the dijet cross section down to $x_A^{\rm obs} \sim 10^{-4}$ (down to $x_A^
  {\rm obs} \sim 10^{-5}$ at the FCC). The ratio of the dijet cross sections on a nucleus
  and the proton, $\sigma_A/(A\sigma_p)$, depends on $x_A^{\rm obs}$ in a similar way as
  the ratio of gluon densities, $g_A(x_A,\mu^2)/[A g_p(x_A,\mu^2)]$, for which current
  nPDFs predict a strong suppression due to nuclear shadowing in the region $x_A^{\rm obs}
  < 0.01$. Dijet photoproduction at future lepton-nucleus colliders can therefore be used
  to test this prediction and considerably reduce the current uncertainties of nPDFs.
\end{abstract}

\preprint{MS-TP-20-14}

\maketitle

\section{Introduction}
\label{sec:intro}

Lepton-nucleus scattering at high energies has traditionally been a fruitful way to
access and study the structure of nuclei in quantum chromodynamics (QCD). Despite numerous
successes and insights, there is an overarching need to continue these studies at
progressively higher energies using colliders. While the plans to use nuclear beams in the
HERA collider at DESY~\cite{Arneodo:1996qa} have not materialized, a
high-energy polarized lepton-proton and lepton-nucleus collider at Brookhaven National
Laboratory (BNL)~\cite{Boer:2011fh,Accardi:2012qut} -- an Electron Ion Collider (EIC) --
has recently been approved.
Further down the road one envisions that the  Large Hadron Collider (LHC) at CERN will be
complemented by a Large Hadron Electron Collider (LHeC) and its higher-energy upgrade
(HE-LHeC)~\cite{AbelleiraFernandez:2012cc,Bruning:2019scy} as well as a Future Circular
Collider (FCC)~\cite{Abada:2019lih}.

The core of the physics program at the future lepton-nucleus colliders is comprised of
deep inelastic scattering (DIS) allowing one to map out various parton 
distributions in nuclei with high precision; 
see, e.g.~Refs.~\cite{Paukkunen:2017phq,Aschenauer:2017oxs,AbdulKhalek:2019mzd,Ethier:2020way}.
 In addition, 
 as one learned from HERA,
photoproduction of jets~\cite{Adloff:2003nr,Abramowicz:2012jz} and
dijets~\cite{Adloff:2002au,Abramowicz:2010cka}
provides useful complimentary information on the QCD (and in particular gluon) structure
of hadrons. This has recently been exploited at the LHC, where ultraperipheral collisions
(UPCs) of heavy ions give an opportunity to study photon-nucleus scattering at
unprecedentedly high energies~\cite{Baltz:2007kq}. In particular, it was shown that
inclusive dijet photoproduction in Pb-Pb UPCs at the LHC can help to reduce the existing
uncertainty in nuclear parton distribution functions (nPDFs) at small $x$ by approximately
a factor of 2~\cite{Guzey:2018dlm,Guzey:2019kik}.

In this work, we calculate the cross section of inclusive dijet photoproduction in
electron-nucleus scattering in the kinematics of the future EIC, LHeC, HE-LHeC, and FCC
using the formalism of collinear factorization, next-to-leading order (NLO) perturbative
QCD, and nCTEQ15~\cite{Kovarik:2015cma} and EPPS16~\cite{Eskola:2016oht} nPDFs. We make
predictions for the cross section distributions as functions of the dijet average
transverse momentum ${\bar p}_T$, the average rapidity $\bar{\eta}$, the observed nuclear
momentum fraction $x_A^{\rm obs}$, and the observed photon momentum fraction
$x_{\gamma}^{\rm obs}$. We compare the kinematic reaches of the four colliders and find
that an increase of the collision energy from the EIC to the LHeC and beyond extends the
coverage in all four considered variables. Notably, the LHeC and HE-LHeC will allow one to
probe the dijet cross section down to $x_A^{\rm obs} \sim 10^{-4}$ (down to $x_A^{\rm obs}
\sim 10^{-5}$ at the FCC), which is two (three) orders of magnitude smaller than that at
the EIC. We then discuss in detail the implications of future measurements of dijet
photoproduction in lepton-nucleus scattering on the determination of nPDFs.

This work continues and extends the analysis of Ref.~\cite{Klasen:2018gtb} by making predictions for high-energy
lepton-nucleus colliders including LHeC, HE-LHeC, and FCC, comparing them to the case of the EIC, and analyzing relative merits
of the four considered colliders.

The remainder of the paper is structured as follows. In Sec.~\ref{sec:formalism}, we recap
the formalism and the input for the calculation of inclusive dijet photoproduction in NLO
perturbative QCD. Our results and their discussion are presented in Sec.~\ref{sec:results}.
A summary of our results is given in Sec.~\ref{sec:conclusions}.

\section{Dijet photoproduction in next-to-leading order QCD}
\label{sec:formalism}

In the framework of collinear factorization and next-to-leading order (NLO) perturbative
QCD~\cite{Klasen:1995ab,Klasen:1996it,Klasen:1997br,Klasen:2011ax,Klasen:2002xb}, the
cross section of dijet photoproduction in $eA\to e+{\rm 2 jets}+X$ electron-nucleus
scattering reads
\begin{equation}
  d\sigma(eA \to e+{\rm 2 jets}+X)=\sum_{a,b} \int dy \int dx_{\gamma} \int dx_A
  f_{\gamma/e}(y)f_{a/\gamma}(x_{\gamma},\mu^2)f_{b/B}(x_A,\mu^2) d\hat{\sigma}
  (ab \to {\rm jets})\,,
 \label{eq:cs}
\end{equation}
where $a,b$ are parton flavors; $f_{\gamma/e}(y)$ is the flux of equivalent photons of the
electron, which depends on the photon light-cone momentum fraction $y$; $f_{a/\gamma}
(x_{\gamma},\mu^2)$ is the PDF of the photon for the resolved photon case (see below),
which depends on the momentum fraction $x_{\gamma}$ and the factorization scale $\mu$;
$f_{b/B}(x_A,\mu^2)$ is the nuclear PDF with $x_A$ being the corresponding parton momentum
fraction; and $d\hat{\sigma}(ab \to {\rm jets})$ is the elementary cross section for the
production of two-parton and three-parton final states emerging as jets in hard scattering
of partons $a$ and $b$. 

The dijet cross section in Eq.~(\ref{eq:cs}) receives two types of contributions: the
resolved photon contribution, when the photon interacts with target partons through its
quark-gluon structure expressed by $f_{a/\gamma}(x_{\gamma},\mu^2)$, and the direct photon
contribution, when the photon enters directly the hard scattering cross section
$d\hat{\sigma}(ab \to {\rm jets})$. At leading-order (LO), the direct photon contribution
has the support exactly at $x_{\gamma} = 1$ and $f_{\gamma/\gamma}(x_{\gamma},\mu^2)=
\delta(1-x_{\gamma})$. At NLO, the separation between the resolved and direct photon
contributions depends on the factorization scheme and scale $\mu$. Indeed, by calculating the
virtual and real corrections to the matrix elements of interest using massless quarks in
dimensional regularization, one can explicitly show that ultraviolet (UV) divergences are
renormalized in the $\overline{\rm MS}$ scheme and infrared (IR) divergences are canceled
and factorized into the nucleus (proton) and photon PDFs, respectively; see
Ref.~\cite{Klasen:2002xb}. For the latter, this can imply a transformation from the
DIS$_\gamma$ to the $\overline{\rm MS}$ scheme. As a result, the direct photon
contribution becomes sizable and in practice dominates the cross section at $x_{\gamma}
\approx 1$ even at NLO.

In our analysis, we used for the photon flux of the electron the improved expression
derived in the Weizs\"acker-Williams approximation~\cite{Frixione:1993yw}
\begin{equation}
 f_{\gamma/e}(y)=\frac{\alpha}{2 \pi} \left[\frac{1+(1-y)^2}{y} \ln \frac{Q^2_{\rm max}(1-y)}{m_e^2 y^2}+2 m_e^2 y
 \left(\frac{1}{Q^2_{\rm max}}-\frac{1-y}{m_e^2 y^2} \right) \right] \,,
 \label{eq:flux}
\end{equation}
where $\alpha$ is the fine-structure constant; $m_e$ is the electron mass; and
$Q^2_{\rm max}$ is the maximal photon virtuality. Motivated by studies of jet
photoproduction at HERA, we take $Q^2_{\rm max}=0.1$ GeV$^2$ and assume that the
inelasticity spans the range of $0 < y < 1$.

For the photon PDFs, we used the GRV HO parametrization~\cite{Gluck:1991jc}, which we
transformed as explained above. These photon PDFs have been tested thoroughly at HERA and
the Large Electron Positron (LEP) collider at CERN and are very robust, especially at high
$x_{\gamma}$ (dominated by the pQCD photon-quark splitting), which is correlated with the
low-$x_A$ region that is of particular interest for this work. For the nuclear PDFs
$f_{b/B}(x_A,\mu^2)$, we employed the nCTEQ15~\cite{Kovarik:2015cma} and
EPPS16~\cite{Eskola:2016oht} parametrizations including both central and error PDFs. The
latter are used to evaluate the theoretical uncertainty bands of our predictions.

\section{Predictions for dijet photoproduction cross sections at future electron-ion colliders}
\label{sec:results}

We performed perturbative NLO QCD calculations of the dijet photoproduction cross section
using Eq.~(\ref{eq:cs}), which was numerically implemented in an NLO parton-level Monte
Carlo~\cite{Klasen:1995ab,Klasen:1996it,Klasen:1997br,Klasen:2011ax,Klasen:2002xb}. This
framework has been successfully tested to describe the HERA and LEP data on dijet
photoproduction on the proton. It implements the anti-$k_T$ algorithm
\cite{Cacciari:2008gp} with a jet radius of $R=0.4$ (we have at most two partons in the
jet) and the following generic conditions on final-state jets: The leading jet has
$p_{T,1} > 5$ GeV, while the other jets have a lower cut on $p_{T,i\neq1}> 4.5$ GeV to
avoid an enhanced sensitivity to soft radiation in the calculated cross
section~\cite{Klasen:1995xe}; all jets have rapidities $|\eta_{1,2}| < 4$. The studied
energy configurations of future electron-ion colliders are summarized in
Table~\ref{table:energy}, where $E_e$ and $E_A$ refer to the electron and nucleus beam
energies, respectively, and $\sqrt{s}$ is the center-of-mass collision energy per
nucleon.
\begin{table}[t]
\caption{Energy configurations of electron-ion colliders considered in this work.}
\begin{center}
\begin{tabular}{|c|c|c|c|}
\hline 
& $E_e$ (GeV) & $E_A$ (TeV) & $\sqrt{s}$ (GeV) \\
\hline
EIC     & 21 & 0.1  &  92 \\
LHeC    & 60 & 2.76 &  812 \\
HE-LHeC & 60 & 4.93 &  1,088 \\
FCC     & 60 & 19.7 &  2,174 \\
\hline
\end{tabular}
\end{center}
\label{table:energy}
\end{table}

In general, i.e., beyond leading order (LO) perturbative QCD, the light-cone momentum
fractions $x_{\gamma}$ and $x_A$ in Eq.~(\ref{eq:cs}) are not directly measurable. Instead
one usually introduces their estimates, which can be defined using the two highest
transverse-energy jets,
\begin{eqnarray}
 x_{\gamma}^{\rm obs}&=&\frac{p_{T,1}e^{-\eta_1}+p_{T,2}e^{-\eta_2}}{2yE_e} \,,\\
 x_{A}^{\rm obs}&=&\frac{p_{T,1}e^{\eta_1}+p_{T,2}e^{\eta_2}}{2E_A} \,,
\end{eqnarray}
where $p_{T,1,2}$  and $\eta_{1,2}$ are the transverse energies and rapidities of the two
jets ($p_{T,1} > p_{T,2}$).

\begin{figure}[t]
\begin{center}
\epsfig{file=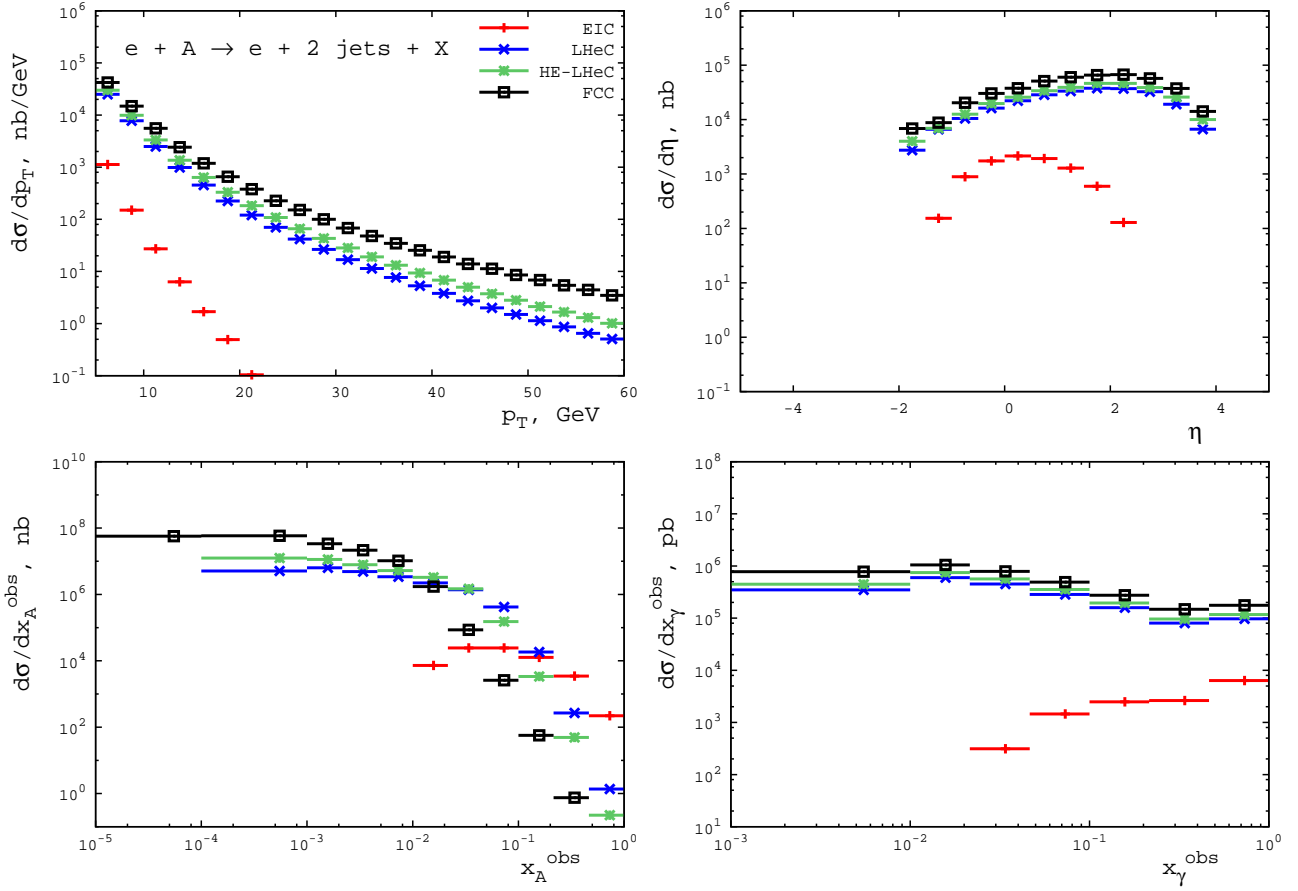,scale=0.75}
\caption{NLO QCD predictions for the dijet photoproduction cross section in $eA\to e+{\rm 2 jets}+X$ electron--nucleus scattering at the EIC, LHeC, HE-LHeC, and FCC as a function of the average dijet transverse momentum ${\bar p}_T$, 
the average rapidity $\bar{\eta}$, and the momentum fractions
$x_A^{\rm obs}$ and $x_{\gamma}^{\rm obs}$. The calculation uses nCTEQ15 nPDFs.}
\label{fig:summary_rates}
\end{center}
\end{figure}

Figure~\ref{fig:summary_rates} summarizes our predictions for the dijet cross section,
Eq.~(\ref{eq:cs}), as a function of the dijet average transverse momentum
${\bar p}_T=(p_{T,1}+p_{T,2})/2$, the average rapidity $\bar{\eta}=(\eta_1+\eta_2)/2$, and
the momentum fractions $x_A^{\rm obs}$ and $x_{\gamma}^{\rm obs}$. The calculations are
performed using the central value of the nCTEQ15 nPDFs. On a logarithmic $y$-scale, EPPS16
nPDFs give indistinguishable results. We find sizable yields in all four considered
variables. In particular, at the EIC the kinematic coverage spans
$5 \leq {\bar p}_T \leq 20$ GeV, $-2 < \bar{\eta} \leq 3$,
$0.03 \leq x_{\gamma}^{\rm obs} \leq 1$, and $0.01 \leq x_A^{\rm obs} \leq 1$;
see also Ref.~\cite{Klasen:2018gtb}. Comparing the kinematic reaches of the four colliders, one
can see from the figure that an increase of the collision energy dramatically expands the
kinematic coverage. At the LHeC, HE-LHeC, and FCC, one probes the dijet cross cross section
in the wider ranges of $5 \leq {\bar p}_T \leq 60$ GeV, $-2 \leq \bar{\eta} \leq 4$,
$10^{-3} \leq x_{\gamma}^{\rm obs} \leq 1$, and $10^{-4} \leq x_A^{\rm obs} \leq 1$ (LHeC
and HE-LHeC), and even $10^{-5} \leq x_A^{\rm obs} \leq 1$ (FCC). 

\begin{figure}[t]
\begin{center}
\epsfig{file=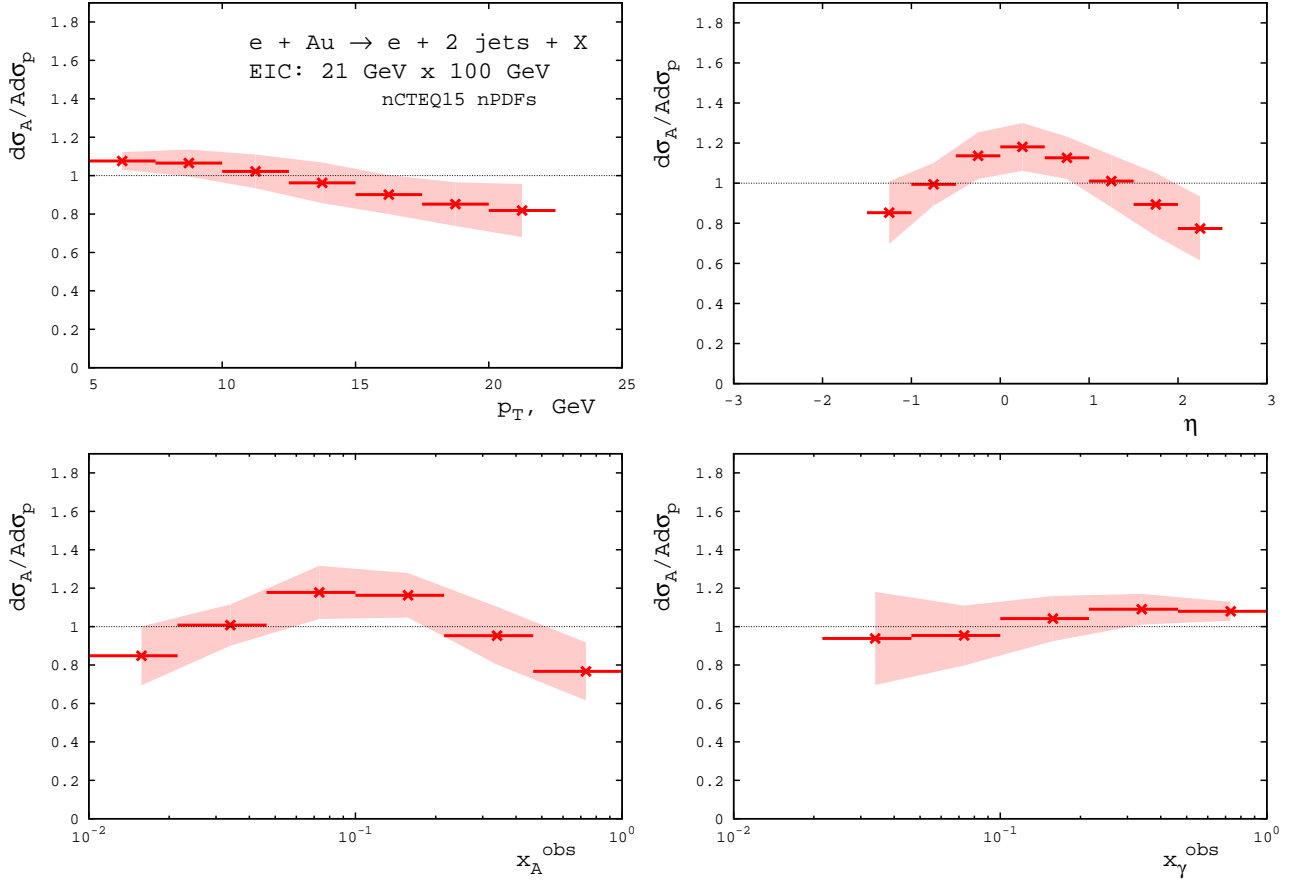,scale=0.75}
\caption{NLO QCD predictions for the ratio of the cross sections of dijet photoproduction on nuclei and the proton as a function of ${\bar p}_T$, 
$\bar{\eta}$, $x_A^{\rm obs}$, and $x_{\gamma}^{\rm obs}$ in the EIC kinematics. The calculation uses central values of
nCTEQ15 nPDFs (solid lines) and 32 sets of error PDFs (shaded band).}
\label{fig:eic_rat_final}
\end{center}
\end{figure}

\begin{figure}[t]
\begin{center}
\epsfig{file=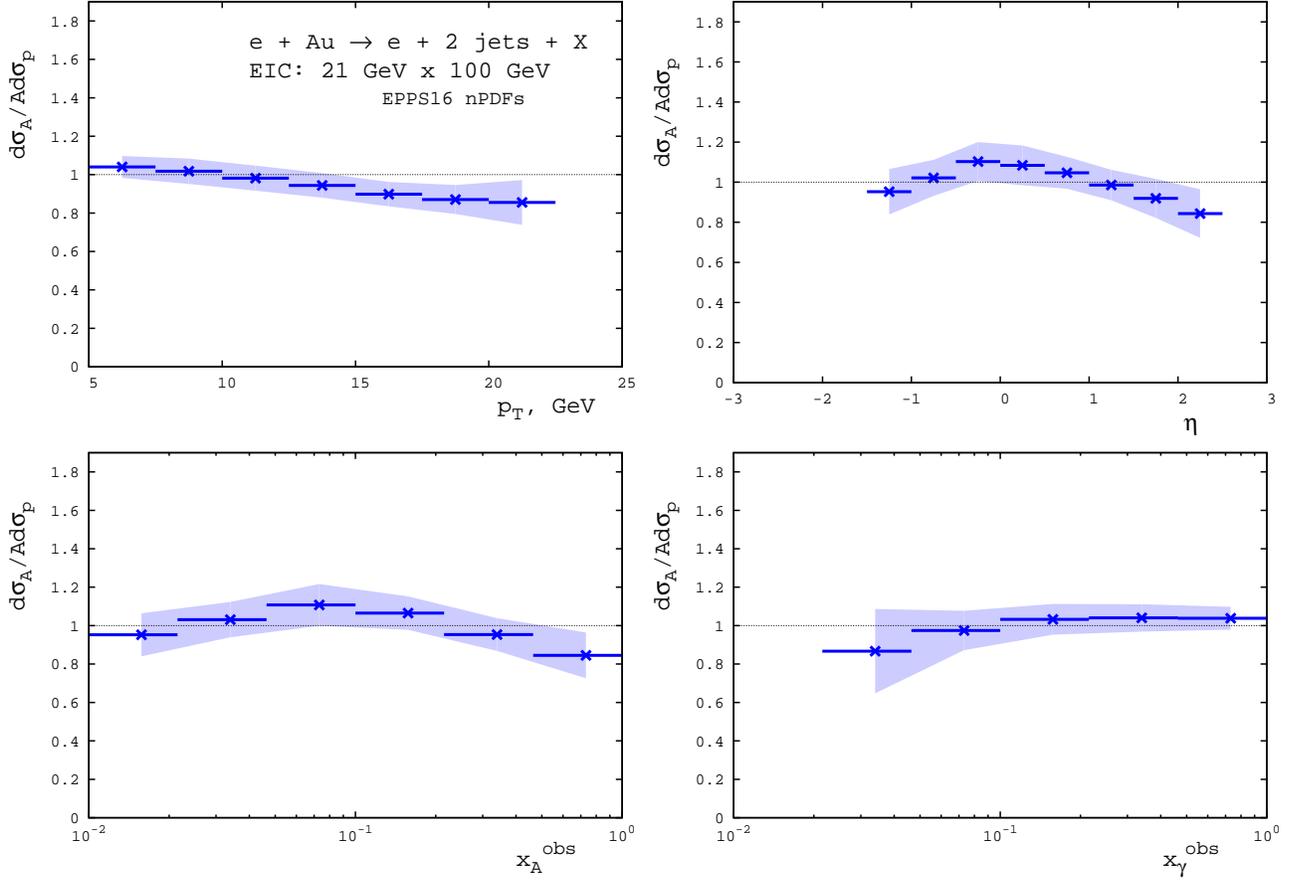,scale=0.75}
\caption{NLO QCD predictions for the ratio of the cross sections of dijet photoproduction on nuclei and the proton as a function of ${\bar p}_T$, 
$\bar{\eta}$, $x_A^{\rm obs}$, and $x_{\gamma}^{\rm obs}$ in the EIC kinematics. The calculation uses central values of 
EPPS16 nPDFs (solid lines) and 40 sets of error PDFs (shaded band).}
\label{fig:eic_rat_final_epps16}
\end{center}
\end{figure}

\begin{figure}[t]
\begin{center}
\epsfig{file=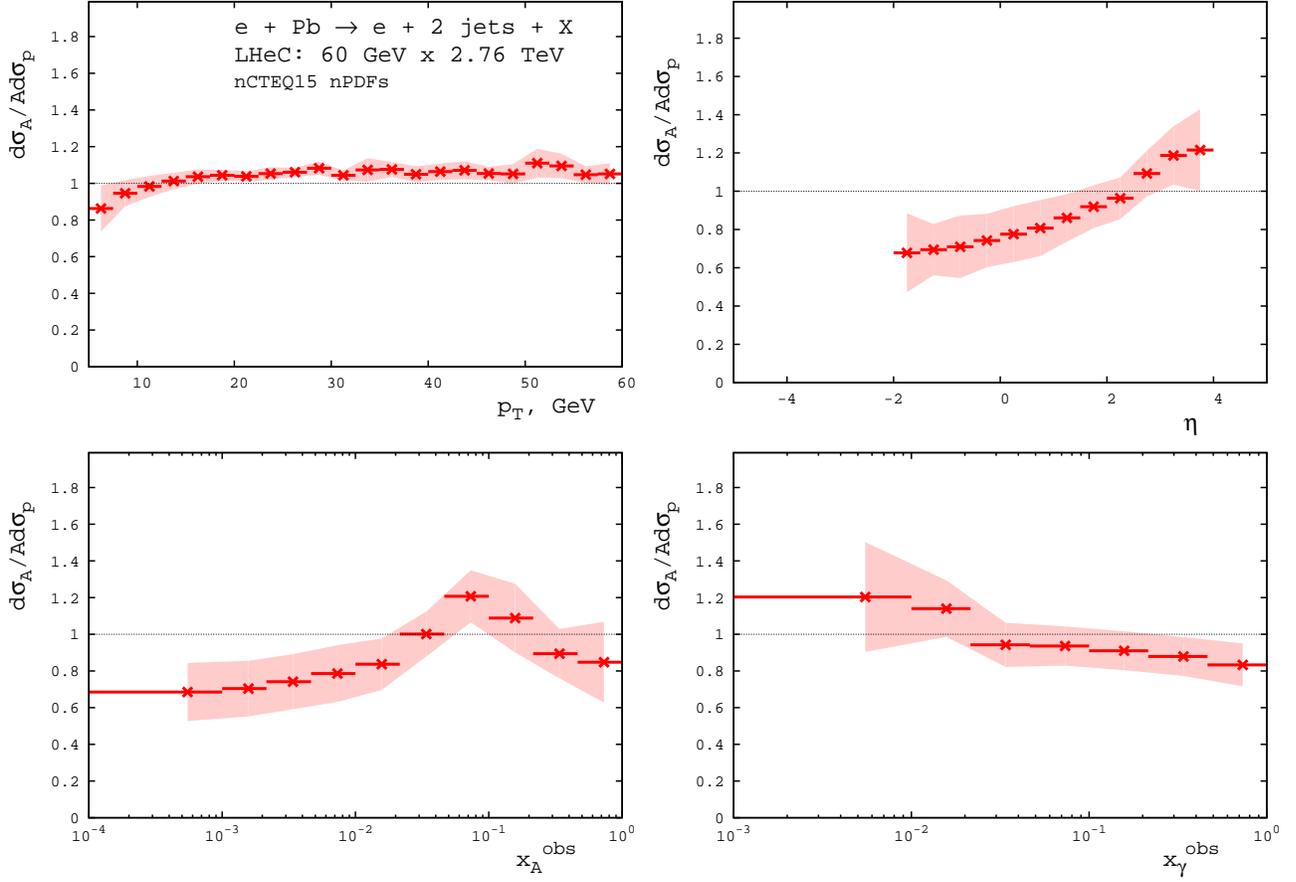,scale=0.75}
\caption{Same as in Fig.~\ref{fig:eic_rat_final} in the LHeC kinematics.}
\label{fig:lhec_rat_final}
\end{center}
\end{figure}

\begin{figure}[t]
\begin{center}
\epsfig{file=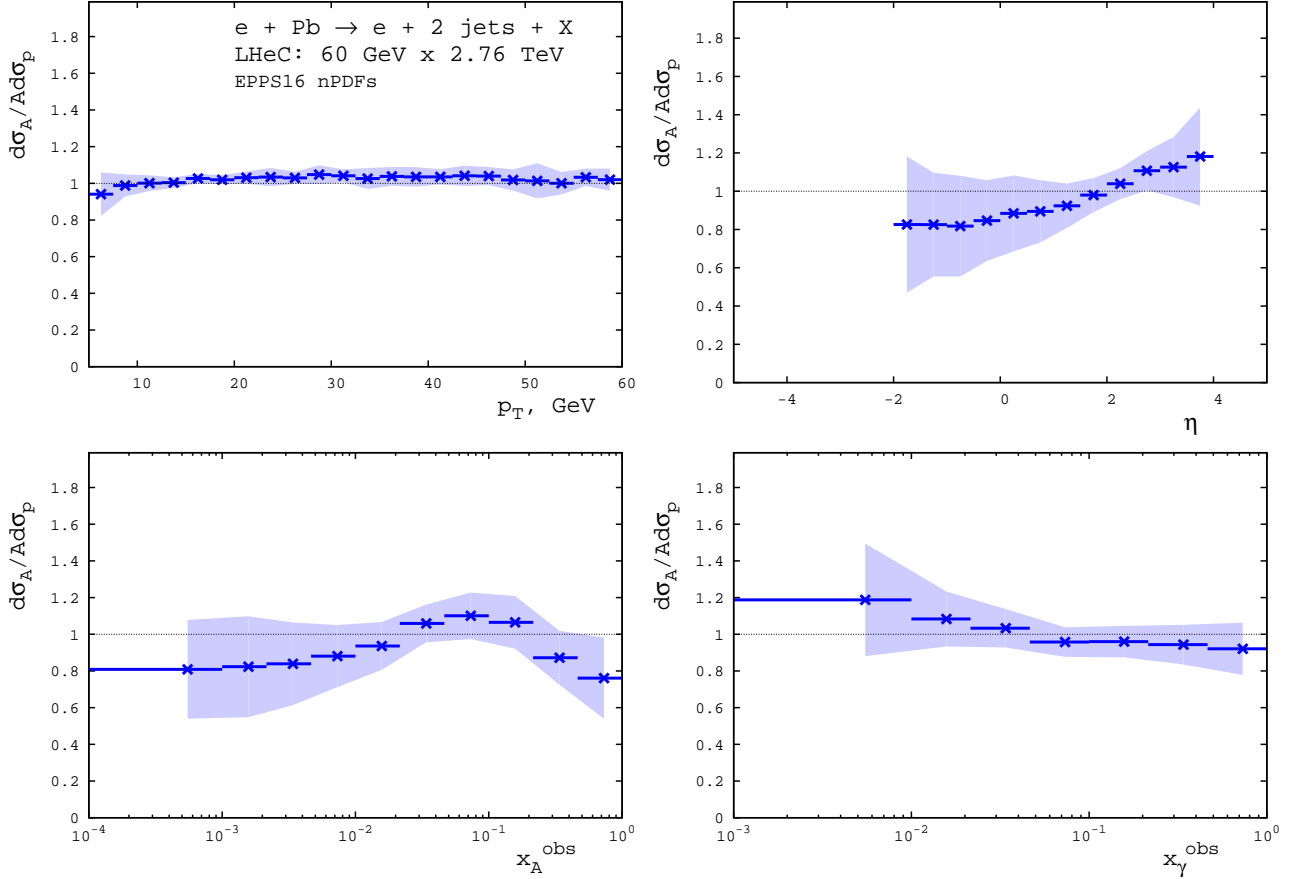,scale=0.75}
\caption{Same as in Fig.~\ref{fig:eic_rat_final_epps16} in the LHeC kinematics.}
\label{fig:lhec_rat_final_epps16}
\end{center}
\end{figure}

To quantify the magnitude of nuclear modifications of the calculated cross section, we show
the ratios of the nuclear cross section, Eq.~(\ref{eq:cs}), to the cross section of dijet
photoproduction on the proton, $d\sigma_A/(A d\sigma_p)$, in Figs.~\ref{fig:eic_rat_final} and \ref{fig:eic_rat_final_epps16}
in the
EIC kinematics and in Figs.~\ref{fig:lhec_rat_final} and \ref{fig:lhec_rat_final_epps16} in the LHeC kinematics. 
The results for the HE-LHeC and FCC
closely resemble those for the LHeC. The cross section ratios are shown as functions of
${\bar p}_T$, $\bar{\eta}$, $x_A^{\rm obs}$, and $x_{\gamma}^{\rm obs}$. In each bin, the
solid lines correspond to the corresponding central value of nPDFs in the calculation of
$d\sigma_A$ and $d\sigma_p$; the shaded band shows the
theoretical uncertainty, which has been  calculated using 32 nCTEQ15 error PDFs~\cite{Kovarik:2015cma} and 40 EPPS16 
error PDF sets~\cite{Eskola:2016oht}. 

In these figures, the results of the calculation using the central value of nPDFs exhibit a
clear nuclear dependence of the presented distributions. At the EIC, the magnitude of
nuclear modifications of the dijet cross section is of the order of $10-20$\% and is
compatible to 
the theoretical uncertainty due to current uncertainties of nCTEQ15 and EPPS16 nPDFs. At the
same time, nuclear modifications of $d\sigma_A/(A d\sigma_p)$ are more pronounced in
the kinematics of LHeC (HE-LHeC, FCC) 
 so that the predicted nuclear suppression of the
$\bar{\eta}$ and $x_{A}^{\rm obs}$ distributions is 
somewhat larger (in the nCTEQ15 case)
than the uncertainty band due 
to nPDFs.  

From the point of view of constraining nPDFs at small $x$, the distribution in
$x_{A}^{\rm obs}$ is the most important one. The shape of $d\sigma_A/(A d\sigma_p)$
repeats that of the ratio of 
the nucleus and proton structure functions $F_{2A}(x,\mu^2)/[A F_{2p}(x,\mu^2)]$ and parton distributions
 $f^{j}_A(x,\mu^2)/[A f^j_p(x,\mu^2)]$ (in particular, the ratio of the nucleus and proton gluon distributions):
the nuclear suppression (shadowing) for
$x_{A}^{\rm obs} < 0.05$ is followed by some enhancement (antishadowing) around
$x_{A}^{\rm obs} \approx 0.1$, which is then followed by the EMC-effect-like suppression
for $x_{A}^{\rm obs} > 0.2$. While the EIC allows one to probe the dijet cross section
down to $x_A^{\rm obs} \approx 0.01$, the LHeC extends the small-$x$ range down to
$x_A^{\rm obs} \approx 10^{-4}$ (down to $x_A^{\rm obs} \approx 10^{-5}$ at FCC). It significantly
enhances the sensitivity to nuclear modifications of nPDFs at small $x$.

An inspection of Figs.~\ref{fig:summary_rates}--\ref{fig:lhec_rat_final_epps16} allows one to qualitatively explain the obtained results. At the
EIC, the dijet cross section is peaked around $x_{A}^{\rm obs} \approx 0.1$, where nPDFs are somewhat enhanced compared to the free proton case,
and, hence,
one expects that $d\sigma_A/(A d\sigma_p) \geq 1$ in the dominant part of the phase space,
and in particular at small $\bar{p}_T$ and large $x_{\gamma}$. It also reveals the
anti-correlation of $x_{\gamma}$ with $x_A$: $d\sigma_A/(A d\sigma_p)$ is simultaneously
enhanced around $x_{A}^{\rm obs} \approx 0.1$ (which corresponds to small $x_A$ in the EIC
kinematics) and for large values of $x_{\gamma}^{\rm obs}$.

At the LHeC, the dijet cross section is dominated by small $x_A$, $x_{A}^{\rm obs}<0.01$.
Hence, one expects that the $d\sigma_A/(A d\sigma_p)$ cross section ratio is suppressed in
most of the phase space, which is indeed observed in Figs.~\ref{fig:lhec_rat_final} and \ref{fig:lhec_rat_final_epps16}. 
The
anti-correlation of $x_{\gamma}$ with $x_A$ is also clearly seen:
$d\sigma_A/(A d\sigma_p) < 1$ for small $x_{A}^{\rm obs}$ and large $x_{\gamma}^{\rm obs}$.

Note that the expected statistical uncertainty of measurements of the cross section of dijet photoproduction  
will be much smaller than the theoretical error bands due to nPDFs shown in Figs.~\ref{fig:eic_rat_final}--\ref{fig:lhec_rat_final_epps16}. 
Indeed, using the projected integrated luminosity of $\int dt \, {\cal L} =10$ fb$^{-1}$/A for all four considered colliders~\cite{Bruning:2019scy,Aschenauer:2017oxs}, one can readily estimate that the expected statistic uncertainty
in each bin in Figs~\ref{fig:summary_rates}-\ref{fig:lhec_rat_final_epps16} should be better than 1\%. The expected systematic uncertainty is expected to be at the level of
2\%; see Refs.~\cite{Klasen:2018gtb,Aschenauer:2017oxs}.
Hence, dijet photoproduction at future lepton-nucleus colliders can be used
  to considerably reduce the current uncertainties of nPDFs.

\section{Conclusions}
\label{sec:conclusions}

We calculated the cross section of inclusive dijet photoproduction in electron-nucleus
scattering in the kinematics of such future lepton-nucleus colliders as EIC, LHeC,
HE-LHeC, and FCC using NLO perturbative QCD and nCTEQ15 and EPPS16 nPDFs. We made
predictions for the cross section distributions as functions of the dijet average
transverse momentum ${\bar p}_T$, the average rapidity $\bar{\eta}$, the nuclear momentum
fraction $x_A^{\rm obs}$, and the photon momentum fraction $x_{\gamma}^{\rm obs}$ and
compared the kinematic reaches of the four colliders.  We found that an increase of the
collision energy from the EIC to the LHeC and beyond extends the coverage in all four
considered variables. Notably, the LHeC and HE-LHeC will allow one to probe the dijet
cross section down to $x_A^{\rm obs} \sim 10^{-4}$ (down to $x_A^{\rm obs} \sim 10^{-5}$
at the FCC). We also calculated the ratio of the dijet cross sections on a nucleus and the
proton, $\sigma_A/(A\sigma_p)$, and showed that it exhibits clear nuclear modifications. 
We found that in the important case of the $x_A^{\rm obs}$ dependence, the shape of
$\sigma_A/(A\sigma_p)$ repeats that 
of the ratio of the nucleus and proton parton distributions and, in particular, the 
$g_A(x,\mu^2)/[A g_p(x,\mu^2)]$ ratio, 
and reveals a
strong suppression due to nuclear shadowing for $x_A^{\rm obs} < 0.01$. This indicates that
dijet photoproduction in lepton-nucleus scattering in the kinematics of the future
lepton-nucleus colliders will be very beneficial to reduce current uncertainties of nPDFs.

\acknowledgments

M.K.~would like to thank the Petersburg Nuclear Physics Institute (PNPI), Gatchina, for the kind
hospitality extended to him during his research visit.
V.G.~would like to thank the Institut f\"ur Theoretische Physik, Westf\"alische
Wilhelms-Universit\"at M\"unster for hospitality. V.G.'s research is supported in part by 
RFBR, Research Project No.~17-52-12070. The authors gratefully acknowledge financial 
support of DFG through Grant No.~KL 1266/9-1 within the framework of the joint
German-Russian project ``New constraints on nuclear parton distribution functions at
small $x$ from dijet production in $\gamma A$ collisions at the LHC".

\end{document}